\documentclass[prb,twocolumn, showpacs]{revtex4}
\usepackage{amsmath,amsthm,amssymb}
\usepackage{latexsym}
\usepackage{array}
\usepackage{amscd,graphicx}
\usepackage{bm,textcomp}

\begin{document}

\title{Some symmetry properties of spin currents and spin polarizations in
multi-terminal mesoscopic spin-orbit coupled systems}
\author{Yongjin Jiang}
\affiliation{Department of physics, Zhejiang Normal University, Jinhua, Zhejiang 321004,
P. R. China}
\author{ Liangbin Hu}
\affiliation{Department of physics and Laboratory of photonic information technology,
South China Normal University, Guangdong 510631, P. R. China }

\begin{abstract}
We study theoretically some symmetry properties of spin currents and spin
polarizations in multi-terminal mesoscopic spin-orbit coupled systems. Based
on a scattering wave function approach, we show rigorously that in the
equilibrium state no finite spin polarizations can exist in a multi-terminal
mesoscopic spin-orbit coupled system ( both in the leads and in the
spin-orbit coupled region ) and also no finite equilibrium terminal spin
currents can exist. By use of a typical two-terminal mesoscopic spin-orbit
coupled system as the example, we show explicitly that the nonequilibrium
terminal spin currents in a multi-terminal mesoscopic spin-orbit coupled
system are non-conservative in general. This non-conservation of terminal
spin currents is not caused by the use of an improper definition of spin
current but is intrinsic to spin-dependent transports in mesoscopic
spin-orbit coupled systems. We also show that the nonequilibrium lateral
edge spin accumulation induced by a longitudinal charge current in a thin
strip of \textit{finite} length of a two-dimensional electronic system with
intrinsic spin-orbit coupling may be non-antisymmetric in general, which
implies that some cautions may need to be taken when attributing the
occurrence of nonequilibrium lateral edge spin accumulation induced by a
longitudinal charge current in such a system to an intrinsic spin Hall
effect.
\end{abstract}

\pacs{72.25.-b,  73.23.-b, 75.47.-m}
\maketitle


\section{Introduction}

The efficient generation of finite spin polarizations and/or spin-polarized
currents in paramagnetic semiconductors by all-electrical means is one of
the principal challenges in semiconductor spin-based electronics.\cite%
{Dassarma2004,semibook2002} For this purpose several interesting ideas have
been proposed based on the spin-orbit ( SO ) interaction character of
electrons in some semiconductor systems.\cite%
{Das90,Hirsch1999,MurakamiScience2003,SinovaPRL2004} One such interesting
idea is the so-called \textit{intrinsic spin Hall effect},\cite%
{MurakamiScience2003,SinovaPRL2004} which is the generation of a finite spin
current perpendicular to an applied charge current in a paramagnetic
semiconductor with intrinsic SO coupling. Such ideas have attracted much
theoretical interests recently$^{7-24}$ and substantial achievements do have
been obtained along these lines,\cite{KatoScience2004,WunderlichPRL2005}
while at the same time they also raised a lot of debates and controversies.%
\cite{RashbaPRB2003,InouePRB2004,MishchenkoPRL2004} A central problem
related to these debates and controversies is that, what is the correct
definition of spin current in a system with strong SO coupling and what is
the actual relation between the spin current and the induced
spin-accumulation in such a system.\cite{Jsinova2005} In most recent studies
the conventional ( \textit{standard} ) definition of spin current ( i.e.,
the expectation value of the product of spin and velocity operators ) have
been applied. As is well know, this conventional definition of spin current
can describe properly spin-polarized transport in a system without intrinsic
SO coupling. However, since spin is not a conserved quantity in a system
with intrinsic SO coupling, the physical meanings of spin current calculated
based on the conventional definition are somewhat ambiguous and the actual
relations between the spin current and the induced spin-accumulation are not
much clear. In fact, as has been noticed in several recent papers,\cite%
{Murakami04,Sun05,shijunren,Jin05} there may exist some serious problems
with this conventional definition of spin current when using it to describe
spin-polarized transport in a system with intrinsic SO coupling. In order to
avoid such serious problems, several alternative definitions of spin current
were proposed in these papers based on different theoretical considerations,
which are significantly different from the conventional one and also
significantly different from each other.\cite%
{Murakami04,Sun05,shijunren,Jin05} Another possible way to circumvent this
problem is to study a \textit{mesoscopic} SO coupled system attached to
external leads. If no SO couplings present in the leads or the SO couplings
in the leads are much weak, then the conventional definition of spin current
can be well applied without ambiguities in the leads. Several recent works
have adopted this strategy\cite%
{LShengPRL2005,NikolicPRB2005,Onoda2005,Hankiewicz2005,JLi2005} and some
interesting results were also obtained. Of course, the study of spin
transport in mesoscopic SO coupled systems is not only of theoretical
interest but also might find some practical applications in the design of
spin-based electronic devices.\cite{Das90}

In this paper we study theoretically some interesting problems related to
spin-dependent transports in multi-terminal mesoscopic SO coupled systems.
We focus our study on the symmetry properties of equilibrium and
nonequilibrium spin currents and spin polarizations in such mesoscopic
structures. As is well known, symmetry analysis is usually of great
theoretical importance in the study of many physical phenomena, including
the spin-dependent transport phenomena in SO coupled systems.\cite{zhaifeng}
Based on the analyses of symmetry properties of equilibrium and
nonequilibrium spin currents and spin polarizations in multi-terminal
mesoscopic SO coupled systems, some controversial issues related to
spin-dependent transports in mesoscopic SO coupled systems will be
investigated in some detail in this paper. Some symmetry properties
discussed in this paper might also be helpful for clarifying some
controversial issues encountered in the study of spin-dependent transports
in macroscopic SO coupled systems. The study carried out in this paper is
based on a scattering wave function approach within the framework of the
standard Landauer-B\"{u}ttiker's formalism. From the theoretical points of
view, this scattering wave function approach is in principle exactly
equivalent to the more frequently employed Green's function approach in
literature\cite{data}. The main merit of this scattering wave function
approach is its conceptual simplicity, and due to its conceptual simplicity,
some symmetry properties of equilibrium and nonequilibrium spin currents and
spin polarizations in a multi-terminal mesoscopic SO coupled system can be
more explicitly shown.

The paper is organized as follows: In section II we will first give a brief
introduction of the structure considered and the approach applied. In
section III we will use the approach introduced in section II to investigate
whether there can exist nonvanishing equilibrium spin polarizations or
nonvanishing equilibrium terminal spin currents in a multi-terminal
mesoscopic SO coupled system. In section IV we will study the symmetry
properties of nonequilibrium spin polarizations and nonequilibrium terminal
spin currents in a typical two-terminal mesoscopic structure with both
Rashba and Dresselhaus SO coupling. Finally in Section V a brief summary of
the main conclusions obtained in the paper will be given.

\section{\label{sec:1}Description of the structure and the scattering wave
function approach}

\begin{figure}[tbh]
\includegraphics[width=7cm,height=6cm]{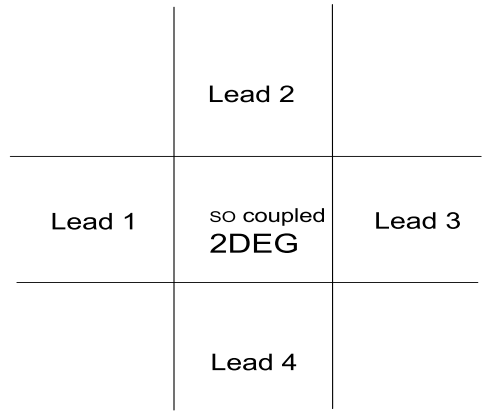}
\caption{Schematic geometry of a multi-terminal mesoscopic SO coupled
system. }
\label{fig:fig1}
\end{figure}

We consider a general multi-terminal mesoscopic structure as shown in Fig.%
\ref{fig:fig1}, where a SO coupled mesoscopic system is attached to several
ideal leads. In a discrete representation, both the SO coupled region and
the ideal leads are described by a tight-binding ( TB ) Hamiltonian, and the
total Hamiltonian for the entire structure reads:
\begin{equation}
\hat{H}=H_{leads}+H_{sys}+H_{s-l}.  \label{eq:1}
\end{equation}%
Here $H_{leads}=\Sigma _{p}H_{p}$ and $H_{p}=-t_{p}\Sigma _{<\mathbf{p}_{i},%
\mathbf{p}_{j}>\sigma }(\hat{C}_{\mathbf{p}_{i}\sigma }^{\dag }\hat{C}_{%
\mathbf{p}_{j}\sigma }+H.C.)$ is the Hamiltonian for an isolated lead $p$,
with $\hat{C}_{\mathbf{p}_{j}\sigma }$ denoting the annihilation operator of
electrons with spin index $\sigma $ at a lattice site $\mathbf{p}_{j}$ in
lead $p$\textit{\ }and\textit{\ }$t_{p}$ the hopping parameter between two
nearest-neighbored lattice sites $\mathbf{p}_{i}$ and $\mathbf{p}_{j}$ in
the lead. We assume that the external leads are ideal and nonmagnetic, i.e.,
no any SO couplings ( or other kinds of spin-flip processes rather than that
induced by the scatterings from the central SO coupled region ) present in
the leads. In such ideal cases the standard definition of spin current can
be well applied without ambiguities in the leads. Usually ( if not specified
) we will choose the $z$ axis ( normal to the 2DEG plane ) as the
quantization axis of spin. $H_{sys}=H_{0}+H_{so}$ is the Hamiltonian for the
isolated SO coupled region, in which $H_{so}$ describes the SO coupling of
electrons and $H_{0}=-t\Sigma _{<\mathbf{r}_{i},\mathbf{r}_{j}>\sigma }(\hat{%
C}_{\mathbf{r}_{i}\sigma }^{\dag }\hat{C}_{\mathbf{r}_{j}\sigma }+H.C.)$
describes the spin-independent hopping of electrons between
nearest-neighbored lattice sites ( denoted by $\langle \mathbf{r}_{i},%
\mathbf{r}_{j}\rangle $ ) in the region. The discrete version of $H_{so}$
will depend on the actual form of the SO interaction, e.g., for the usual
Rashba and k-linear Dresselhaus SO coupling, one has 
\begin{subequations}
\begin{equation}
H_{so}^{R}=-t_{R}\sum_{\mathbf{r}_{i}}[i(\hat{\Psi}_{\mathbf{r}_{i}}^{\dag
}\sigma ^{x}\hat{\Psi}_{\mathbf{r}_{i}+\mathbf{\Delta }_{y}}-\hat{\Psi}_{%
\mathbf{r}_{i}}^{\dag }\sigma ^{y}\hat{\Psi}_{\mathbf{r}_{i}+\mathbf{\Delta }%
_{x}})+H.C.],  \label{2a}
\end{equation}%
\begin{equation}
H_{so}^{D}=-t_{D}\sum_{\mathbf{r}_{i}}[i(\hat{\Psi}_{\mathbf{r}_{i}}^{\dag
}\sigma ^{y}\hat{\Psi}_{\mathbf{r}_{i}+\mathbf{\Delta }_{y}}-\hat{\Psi}_{%
\mathbf{r}_{i}}^{\dag }\sigma ^{x}\hat{\Psi}_{\mathbf{r}_{i}+\mathbf{\Delta }%
_{x}})+H.C.],  \label{2b}
\end{equation}%
where $t_{R}$ and $t_{D}$ are the Rashba and Dresselhaus SO coupling
strength, respectively, $\hat{\Psi}_{\mathbf{r}_{i}}=(\hat{C}_{\mathbf{r}%
_{i},\uparrow },\hat{C}_{\mathbf{r}_{i},\downarrow })$ denotes the spinor
annihilation operators, and $\mathbf{\Delta }_{x}$ and $\mathbf{\Delta }_{y}$
denote the lattice vectors between two nearest-neighbored lattice sites
along the $x$\textit{\ }and\textit{\ }$y$ directions, respectively. The last
term in the Hamiltonian (1) describes the coupling between the leads and the
SO coupled region, $H_{s-l}=-\Sigma _{p}t_{ps}\sum_{n}(\hat{C}_{\mathbf{p}%
_{n}\sigma }^{\dag }\hat{C}_{\mathbf{r}_{n}\sigma }+H.C.)$, where $\mathbf{p}%
_{n}$ denotes a boundary lattice site in lead $p$ connected directly to a
boundary lattice site $\mathbf{r}_{n}$ in the SO coupled region and $t_{ps}$
the hopping parameter between lead $p$ and the SO coupled region. It should
be noticed that in general the TB Hamiltonian will also contain an on-site
energy term, which is not explicitly shown above.

Now we consider the scattering of a conduction electron incident on a lead $%
p $ by the SO coupled region. For conveniences, we will adopt a separate
local coordinate frame in each lead, i.e., we will use a double coordinate
index $(x_{p},y_{p})$ to denote a lattice site in lead $p$, where $%
x_{p}=1,2,...,\infty $ ( away from the border between the lead and the SO
coupled region ) and $y_{p}=1,...,N_{p}$ ( $N_{p}$ is the width of lead $p$
). In the local coordinate frame, the spatial wave function of a conduction
electron incident on lead $p$ will be given by $e^{-ik_{m}^{p}x_{p}}\chi
_{m}^{p}(y_{p})$, where $k_{m}^{p}$ denotes the longitudinal wave vector and
$\chi _{m}^{p}(y_{p})$ the transverse spatial wave function and $m$ the
label of the transverse mode. The longitudinal wave vector will be
determined by the following dispersion relation, $-2t\cos
(k_{m}^{p})+\varepsilon _{m}^{p}=E$, where $\varepsilon _{m}^{p}$ is the
eigen-energy of the $m$th transverse mode and $E$ the energy of the incident
electron. It should be noted that, due to the presence of SO coupling in the
central scattering region, spin-flip processes ( e.g., the spin-flip
reflection ) will be induced in the leads when a conduction electron is
scattered or reflected by the central scattering region, even if the leads
are ideal and nonmagnetic ( which is the just case assumed in the present
paper ). Due to this fact, for a conduction electron incident from the $m$th
transverse channel of lead $p$ and with a \textit{given} spin index $\sigma $%
, both the scattering wave function $|\psi ^{pm\sigma }(\mathbf{r})\rangle $
in the central SO coupled region and the scattering wave function $|\psi
^{pm\sigma }(\mathbf{x}_{p^{\prime }})\rangle $ ( $\mathbf{x}_{p^{\prime
}}\equiv (x_{p^{\prime }},y_{p^{\prime }})$ ) in a lead $p^{\prime }$ will
be inherently a superposition of a spin-up and a spin-down components,
\end{subequations}
\begin{subequations}
\begin{eqnarray}
|\psi ^{pm\sigma }(\mathbf{r})\rangle &=&\sum_{\sigma ^{^{\prime }}}\psi
_{\sigma ^{^{\prime }}}^{pm\sigma }(\mathbf{r})\hat{C}_{\mathbf{r}%
_{i},\sigma ^{\prime }}^{\dag }|0\rangle , \\
|\psi ^{pm\sigma }(\mathbf{x}_{p^{\prime }})\rangle &=&\sum_{\sigma \prime
}\psi _{\sigma \prime }^{pm\sigma }(\mathbf{x}_{p^{\prime }})\hat{C}_{%
\mathbf{x}_{p^{\prime }},\sigma ^{\prime }}^{\dag }|0\rangle ,
\end{eqnarray}%
where $|0\rangle $ stands for the vacuum state. The spin-resolved components
$\psi _{\sigma \prime }^{pm\sigma }(\mathbf{x}_{p^{\prime }})$ of the
scattering wave function in lead $p^{\prime }$ can be expressed in the
following general form,
\end{subequations}
\begin{eqnarray}
\psi _{\sigma \prime }^{pm\sigma }(\mathbf{x}_{p^{\prime }}) &=&\delta
_{pp^{\prime }}\delta _{\sigma \sigma \prime }e^{-ik_{m}^{p}x_{p}}\chi
_{m}^{p}(y_{p})  \notag \\
&&+\sum_{m^{\prime }\in {p^{\prime }}}\phi _{p^{\prime }m^{\prime }\sigma
^{\prime }}^{pm\sigma }e^{ik_{m^{\prime }}^{p^{\prime }}x_{p^{\prime }}}\chi
_{m^{\prime }}^{p^{\prime }}(y_{p^{\prime }}),  \label{eq:swavefunction}
\end{eqnarray}%
where $\phi _{p^{\prime }m^{\prime }\sigma ^{\prime }}^{pm\sigma }$ stands
for the scattering amplitude from the ($m\sigma $) channel of lead $p$ ( the
\textit{incident} mode ) to the ($m^{\prime }\sigma ^{\prime }$) channel of
lead $p^{\prime }$ ( the \textit{out-going} mode ). If $p^{\prime }=p$, the
second term on the right-hand side of Eq.(4) will denote actually the
spin-resolved reflected waves in lead $p$ and $\phi _{p^{\prime }m^{\prime
}\sigma ^{\prime }}^{pm\sigma }$ denote the spin-flip ( $\sigma \neq \sigma
^{^{\prime }}$ ) and non-spin-flip ( $\sigma =\sigma ^{^{\prime }}$ )
reflection amplitudes. The scattering amplitudes $\phi _{p^{\prime
}m^{\prime }\sigma ^{\prime }}^{pm\sigma }$ can be obtained by solving the
Schr\"{o}dinger equation for the entire structure. Since Eq.(\ref%
{eq:swavefunction}) is just a linear combination of all out-going modes with
the same energy $E$ in lead $p^{\prime }$, the Schr\"{o}dinger equation is
satisfied automatically in lead $p^{\prime }$ except at those boundary
lattice sites in lead $p^{\prime }$ connected directly to the SO coupled
region. Due to the coupling between the leads and the SO coupled region, the
amplitudes of the wave function at these boundary lattice sites ( which are
determined by the scattering amplitudes $\phi _{qn\sigma ^{\prime
}}^{pm\sigma }$ ) must be solved simultaneously with the wave function $\psi
^{pm\sigma }(\mathbf{r})$ inside the SO coupled region. From the discrete
version of the Schr\"{o}dinger equation, one can show that the amplitudes of
the wave function in the SO coupled region and at those boundary lattice
sites of the external leads connected directly to the SO coupled region will
satisfy the following coupled equations:%
\begin{subequations}
\begin{eqnarray}
E\psi _{\sigma ^{^{\prime }}}^{pm\sigma }(\mathbf{r}_{s}) &=&\sum_{\mathbf{r}%
_{s}^{^{\prime }}\mathbf{,}\sigma ^{^{\prime \prime }}}H_{sys}(\mathbf{r}%
_{s}\sigma ^{^{\prime }},\mathbf{r}_{s}^{^{\prime }}\sigma ^{^{\prime \prime
}})\psi _{\sigma ^{^{\prime \prime }}}^{pm\sigma }(\mathbf{r}_{s}^{^{\prime
}})  \notag \\
&-&\sum_{p^{\prime },y_{p^{\prime }}}t_{p^{\prime }s}\delta _{\mathbf{r}%
_{s},n_{p^{\prime }y^{\prime }}}\psi _{\sigma ^{^{\prime }}}^{pm\sigma
}(1,y_{p^{\prime }}^{\prime }),
\end{eqnarray}%
\begin{eqnarray}
E\psi _{\sigma ^{^{\prime }}}^{pm\sigma }(\mathbf{x}_{p^{^{\prime
}}}^{^{\prime }}) &=&\sum_{\mathbf{x}_{p^{^{\prime }}}^{\prime \prime
}}H_{p^{^{\prime }}}(\mathbf{x}_{p^{\prime }}^{\prime }\sigma ^{^{\prime }},%
\mathbf{x}_{p^{\prime }}^{\prime \prime }\sigma ^{^{\prime }})\psi _{\sigma
^{\prime }}^{pm\sigma }(\mathbf{x}_{p^{\prime }}^{\prime \prime })  \notag \\
&&-\sum_{\mathbf{r}_{s}}t_{p^{\prime }s}\delta _{\mathbf{r}_{s},n_{p^{\prime
}y^{\prime }}}\psi _{\sigma ^{\prime }}^{pm\sigma }(\mathbf{r}_{s}),
\end{eqnarray}%
where $(\mathbf{r}_{s},\mathbf{r}_{s}^{\prime })$ denote two
nearest-neighbored lattice sites in the SO coupled region and $(\mathbf{x}%
_{p^{\prime }}^{\prime },\mathbf{x}_{p^{\prime }}^{\prime \prime })$ two
nearest-neighbored boundary lattice sites in lead $p^{\prime }$. ( For
simplicity of notation, in the subscript of the Kronecker $\delta -$function
we have used simply a symbol $n_{p^{\prime }y^{\prime }}$ to denote a
boundary lattice site in the SO coupled region which is connected directly
to a boundary lattice site $\mathbf{x}_{p^{\prime }}^{\prime
}=(1,y_{p^{\prime }}^{\prime })$ in lead $p^{\prime }$. ) The matrix
elements $H_{sys}(\mathbf{r}_{s}\sigma ^{^{\prime }},\mathbf{r}%
_{s}^{^{\prime }}\sigma ^{^{\prime \prime }})$ and $H_{p^{^{\prime }}}(%
\mathbf{x}_{p^{\prime }}^{\prime }\sigma ^{^{\prime }},\mathbf{x}_{p^{\prime
}}^{\prime \prime }\sigma ^{^{\prime }})$ can be written down directly from
the Hamiltonian (1). Eq.(5a) and (5b) are the match conditions of the
scattering wave function on the borders between the SO coupled region and
the leads, from which both the scattering wave function in the entire
structure and all scattering amplitudes can be obtained simultaneously. Some
details of the derivations are given in the appendix.

\section{Some rigorous properties of equilibrium states}

A controversial issue encountered in the study of spin-polarized transports
in intrinsically SO coupled systems is that wether there can exist
nonvanishing \textit{equilibrium background spin currents} in such systems.
Recently Rashba pointed out that, in a bulk two-dimensional electron gas (
2DEG ) with Rashba SO coupling, a finite equilibrium background spin current
could be obtained if the conventional definition of spin current is applied
to such systems.\cite{RashbaPRB2003} If this equilibrium background spin
current does exist, it would imply that nonvanishing equilibrium spin
polarizations should also exist near the edges of such a system due to the
flow of the equilibrium background spin current. It was argued in Ref.[8]
that such equilibrium background spin currents are an artefact caused by the
improper use of the conventional definition of spin current to an
intrinsically SO coupled system, i.e., the conventional definition of spin
current cannot be applied in the presence of intrinsic SO coupling. Since it
seemed that no consensus had been arrived on whether there is a uniquely
correct definition for spin current in a SO coupled system\cite%
{Murakami04,Sun05,shijunren,Jin05}, it would be meaningful if this
controversial issue can be investigated in a somewhat different way. In this
section we will use the scattering wave function approach introduced in
section II to investigate wether there can exist nonvanishing equilibrium
background spin currents and/or nonvanishing equilibrium spin polarizations
in a multi-terminal mesoscopic SO coupled system. Based on some simple but
rigorous arguments, we will show that no finite equilibrium spin
polarizations and/or finite equilibrium terminal spin currents can exist in
a multi-terminal mesoscopic SO coupled system.

\subsection{Absence of equilibrium spin polarizations}

In the tight-binding representation, the operator for the local spin density
at a lattice site $i$ reads
\end{subequations}
\begin{equation}
\hat{\vec{S}}(i)=\frac{\hbar }{2}\sum_{\alpha \beta }\hat{C}_{i\alpha
}^{\dagger }\vec{\sigma}_{\alpha \beta }\hat{C}_{i\beta }.  \label{operator}
\end{equation}%
( For simplicity of notation, from now on we will use simply a symbol $i$ to
denote a lattice site in the entire structure, i.e., both in the SO coupled
region and in the external leads. ) Under time reversal transformation, the
local spin density operator will transform as $\hat{\vec{S}}(i)\rightarrow
\hat{T}\hat{\vec{S}}(i)\hat{T}^{-1}=-\hat{\vec{S}}(i)$ and the spin operator
transform as $\vec{\sigma}_{\alpha \beta }\rightarrow \hat{T}\vec{\sigma}%
_{\alpha \beta }\hat{T}^{-1}=(-1)^{\alpha +\beta }\vec{\sigma}_{\bar{\alpha}%
\bar{\beta}}^{\ast }=-\vec{\sigma}_{\alpha \beta }$, where $\bar{\alpha}%
\equiv -\alpha $, $\bar{\beta}\equiv -\beta $, $\hat{T}\equiv i\sigma _{y}%
\hat{K}$ denotes the time-reversal transformation operator and $\hat{K}$ the
conjugate operator.

Within the framework of the standard Landauer-B\"{u}ttiker's formalism, any
physical quantities of a mesoscopic system are contributed to by all
scattering states of conduction electrons incident from all contacts. These
scattering states constitute an ensemble which can be specified by a
chemical potential $\mu _{p}$ for each contact through their separate Fermi
distribution function $f(E,\mu _{p})$. For the problems discussed in the
present paper, this ensemble consists of all scattering states described by
the scattering wave functions $\{\psi ^{pm\sigma }\}$ given by Eqs.(3-4). In
order that there is only one particle feeding into each incident channel\cite%
{data}, when we use this ensemble of scattering wave functions to calculate
the expectation value of an operator, one should first normalize the
scattering wave function $\psi ^{pm\sigma }$ by a factor of $1/\sqrt{L}$ ( $%
L\rightarrow \infty $ is the length of lead $p$ ), corresponding to that one
changes the incident wave functions from $e^{ik_{m}^{p}x_{p}}$ to $%
e^{ik_{m}^{p}x_{p}}/\sqrt{L}$\cite{data}. By use of the ensemble of the
normalized scattering wave functions $\{\psi ^{pm\sigma }\}$ and noticing
that the density of states ( DOS ) for the $m$th transverse mode of lead $p$
is given by $\frac{L}{2\pi }\frac{dk}{dE}=\frac{L}{2\pi \hbar v_{pm}}$,
where $v_{pm}=\frac{2t_{p}}{\hbar }\sin (k_{m}^{p})$ is the longitudinal
velocity of the $m$th transverse mode of lead $p$, then one can see that the
local spin density at a lattice site $i$ ( either in the SO coupled region
or in the external leads ) will be given by
\begin{eqnarray}
\langle \hat{\vec{S}}(i)\rangle &=&\sum_{pm\sigma }\int \frac{{d}E}{2\pi }%
f(E,\mu _{p})\frac{1}{\hbar v_{pm}}  \notag \\
&&\times {\sum_{\alpha ,\beta }}[\psi _{\alpha }^{pm\sigma \ast }(i)(\frac{%
\hbar }{2}\vec{\sigma})_{\alpha \beta }\psi _{\beta }^{pm\sigma }(i)+H.C.],
\label{eq:summationspin}
\end{eqnarray}%
where $\psi ^{pm\sigma }$ is the scattering wave function corresponding to
an incident electron from the $(m\sigma )$ channel of lead $p$ with a given
energy $E$. \textit{This formula is valid both in the equilibrium and in the
nonequilibrium states.} If the system is in an equilibrium state, the
chemical potential $\mu _{p}$ will be independent of the lead label, i.e., $%
\mu _{p}\equiv \mu $ and $f(E,\mu _{p})\equiv f(E,\mu )$. Then in Eq.(7) the
summation $\sum_{pm\sigma }[\ldots ]$ can be performed first before carrying
out the integration over energy $E$, and the result for this summation can
be expressed as
\begin{eqnarray}
&&\sum_{\alpha \beta }\sum\limits_{pm\sigma }[\psi _{\alpha }^{pm\sigma \ast
}(i)(\frac{\hbar }{2}\vec{\sigma})_{\alpha \beta }\psi _{\beta }^{pm\sigma
}(i)/\hbar v_{pm}+H.C.]  \notag \\
&=&\sum_{\alpha \beta }[A_{\beta \alpha }(i,i;E)(\frac{\hbar }{2}\vec{\sigma}%
)_{\alpha \beta }+H.C.]  \notag \\
&=&i\sum_{\alpha \beta }\{[G^{R}(E)-G^{A}(E)]_{i\beta ,i\alpha }(\frac{\hbar
}{2}\vec{\sigma})_{\alpha \beta }-H.C.\}.  \label{eq:sumab}
\end{eqnarray}%
Here $G^{R,A}(E)=[E\mathbf{I-}\hat{H}\pm i0^{+}]^{-1}$ is just the retarded
and advanced Green's functions for the system, whose explicit spin-resolved
matrix forms are given by $G_{i\alpha ,j\beta }^{R,A}(E)=\sum_{p^{\prime
}m^{\prime }\sigma ^{\prime }}\psi _{\alpha }^{p^{\prime }m^{\prime }\sigma
^{\prime }\ast }(i)\psi _{\beta }^{p^{\prime }m^{\prime }\sigma ^{\prime
}}(j)/[E\mathbf{-}E^{^{\prime }}\pm i0^{+}]^{-1}$ ( $E^{^{\prime }}$ is the
incident energy corresponding to a scattering wave function $\psi
^{p^{\prime }m^{\prime }\sigma ^{\prime }}$ ); and $A_{\beta \alpha
}(j,i;E)\equiv \sum\limits_{p^{\prime }m^{\prime }\sigma ^{\prime }}\psi
_{\alpha }^{p^{\prime }m^{\prime }\sigma ^{\prime }\ast }(i)\psi _{\beta
}^{p^{\prime }m^{\prime }\sigma ^{\prime }}(j)/\hbar v_{p^{\prime }m^{\prime
}}=i[G^{R}(E)-G^{A}(E)]_{j\beta ,i\alpha }$ is the spin-resolved spectral
function. If the total Hamiltonian $\hat{H}$ for the entire system is
time-reversal invariant, the retarded and advanced Green's functions can be
related by the time reversal transformation as
\begin{equation}
G_{i\alpha ,j\beta }^{A}=(\hat{T}G^{R}\hat{T}^{-1})_{i\alpha ,j\beta
}=(-1)^{\alpha +\beta }G_{i\bar{\alpha},j\bar{\beta}}^{R\ast }
\label{eq:timegg}
\end{equation}%
Combining Eq.(\ref{eq:sumab}) and Eq.(\ref{eq:timegg}) and taking into
account the fact that $\hat{T}\vec{\sigma}_{\alpha \beta }\hat{T}%
^{-1}=(-1)^{\alpha +\beta }\vec{\sigma}_{\bar{\alpha}\bar{\beta}}^{\ast }=-%
\vec{\sigma}_{\alpha \beta }$, one gets immediately that the right-hand side
of Eq.(\ref{eq:sumab}) should vanish exactly, thus the spin density given by
Eq.(7) vanishes exactly in the equilibrium state, suggesting that no finite
equilibrium spin polarizations can survive at any lattice site $i$ in the
entire structure ( both in the SO coupled region and in the leads ). It
should be noted that in arriving at this conclusion we have only made use of
the assumption that the total Hamiltonian $\hat{H}$ for the entire system is
time-reversal invariant ( which should be the case in the absence of
magnetic fields ) and did not involve the actual form of the SO coupling in
the system, so it is a much general conclusion.

\subsection{Absence of equilibrium terminal spin currents}

In this subsection we discuss whether there can exist nonvanishing
equilibrium terminal spin currents in a multi-terminal mesoscopic SO coupled
system. Since we have assumed that the leads are ideal and nonmagnetic (
i.e., described by a simple Hamiltonian $\hat{H}_{p}=-t_{p}\Sigma _{<\mathbf{%
p}_{i},\mathbf{p}_{j}>\sigma }(\hat{C}_{\mathbf{p}_{i}\sigma }^{\dag }\hat{C}%
_{\mathbf{p}_{j}\sigma }+H.C.)$ ), the conventional definitions of charge
and spin currents can be well applied in the leads without ambiguities.
According to the conventional definitions and in the lattice representation,
the charge current and spin current ( with spin parallel to the $\alpha $
axis\cite{note1} ) flowing from a lattice site $\mathbf{p}_{i}$ to a
nearest-neighbored lattice site $\mathbf{p}_{j}$ in lead $p$ can be given by
the the corresponding particle density current as following,
\begin{subequations}
\begin{equation}
\hat{I}_{p,\mathbf{p}_{i}\rightarrow \mathbf{p}_{j}}=e[\hat{J}_{\mathbf{p}%
_{i}\rightarrow \mathbf{p}_{j}}^{+}+\hat{J}_{\mathbf{p}_{i}\rightarrow
\mathbf{p}_{j}}^{-}],
\end{equation}%
\begin{equation}
\hat{I}_{p,\mathbf{p}_{i}\rightarrow \mathbf{p}_{j}}^{\alpha }=\frac{\hbar }{%
2}[\hat{J}_{\mathbf{p}_{i}\rightarrow \mathbf{p}_{j}}^{+}-\hat{J}_{\mathbf{p}%
_{i}\rightarrow \mathbf{p}_{j}}^{-}],
\end{equation}%
where $\hat{I}_{p,\mathbf{p}_{i}\rightarrow \mathbf{p}_{j}}$ denotes the
charge current operator and $\hat{I}_{p,\mathbf{p}_{i}\rightarrow \mathbf{p}%
_{j}}^{\alpha }$ the spin current operator ( with spin parallel to the $%
\alpha $ axis ) and $\hat{J}_{\mathbf{p}_{i}\rightarrow \mathbf{p}%
_{j}}^{\sigma }$ the spin-resolved particle density current operator and $%
\sigma =\pm $ denotes the spin-up and spin-down states with respect to the $%
\alpha $ axis. From the Heisenberg equation of motion for the on-site
particle density: $\frac{d}{dt}\hat{N}_{\mathbf{p}_{i}}=\frac{1}{i\hbar }[%
\hat{N}_{\mathbf{p}_{i}},\hat{H}_{p}]$, where $\hat{N}_{\mathbf{p}_{i}}=\hat{%
C}_{\mathbf{p}_{i}\sigma }^{\dag }\hat{C}_{\mathbf{p}_{i}\sigma }$ is the
the on-site particle density operator in lead $p$, one can show easily that
the spin-resolved particle density current flowing from a lattice site $%
\mathbf{p}_{i}$ to a nearest-neighbored lattice $\mathbf{p}_{j}$ in lead $p$
will be given by
\end{subequations}
\begin{equation}
\hat{J}_{\mathbf{p}_{i}\rightarrow \mathbf{p}_{j}}^{\sigma }=\frac{it_{p}}{%
\hbar }(\hat{C}_{\mathbf{p}_{j}\sigma }^{\dag }\hat{C}_{\mathbf{p}_{i}\sigma
}-\hat{C}_{\mathbf{p}_{i}\sigma }^{\dag }\hat{C}_{\mathbf{p}_{j}\sigma }).
\end{equation}

Now we calculate the terminal charge and spin currents flowing along the
\textit{longitudinal }direction\textit{\ }of a lead $q$. Firstly we consider
the contribution of an incident electron from the $(m\sigma )$ channel of
lead $p$ to the longitudinal charge and spin currents ( with spin parallel
to the $\alpha $ axis ) flowing through a transverse cross-section ( saying,
e.g., the cross-section at $x=x_{q}$ ) of lead $q$, which by definition will
be given by
\begin{subequations}
\begin{eqnarray}
\langle \hat{I}_{q}\rangle _{pm\sigma } &=&\frac{1}{L}\sum_{y_{q}}\langle
\psi ^{pm\sigma }(x_{q}+1,y_{q})|  \notag \\
&&\times \hat{I}_{q,(x_{q},y_{q})\rightarrow (x_{q}+1,y_{q})}|\psi
^{pm\sigma }(x_{q},y_{q})\rangle   \notag \\
&=&\frac{e}{L}\left\{ \sum_{n,\sigma ^{\prime }}v_{qn}|\phi _{qn\sigma
^{\prime }}^{pm\sigma }|^{2}-v_{pm}\delta _{pq}\right\} ,
\end{eqnarray}%
\begin{eqnarray}
\langle \hat{I}_{q}^{\alpha }\rangle _{pm\sigma } &=&\frac{1}{L}%
\sum_{y_{q}}\langle \psi ^{pm\sigma }(x_{q}+1,y_{q})|  \notag \\
&&\times \hat{I}_{q,(x_{q},y_{q})\rightarrow (x_{q}+1,y_{q})}^{\alpha }|\psi
^{pm\sigma }(x_{q},y_{q})\rangle   \notag \\
&=&\frac{h}{4\pi L}\left\{ \sum_{n}v_{qn}\left[ |\phi _{qn+}^{pm\sigma
}|^{2}-|\phi _{qn-}^{pm\sigma }|^{2}\right] -\sigma v_{pm}\delta
_{pq}\right\} ,  \notag \\
&&
\end{eqnarray}%
where $(x_{q},y_{q})$ and $(x_{q}+1,y_{q})$ denote two nearest-neighbored
lattice sites along the longitudinal direction of lead $q$, $\frac{1}{\sqrt{L%
}}|\psi ^{pm\sigma }(x_{q},y_{q})\rangle $ denotes the \textit{normalized}
scattering wave function in lead $q$ corresponding to the incident electron
from the $(m\sigma )$ channel of lead $p$ ( which is given by Eqs.(3--4) ),
and $v_{qn}=\frac{2t_{p}}{\hbar }\sin (k_{n}^{q})$ denotes the longitudinal
velocity of the $n$th transverse mode in lead $q$ and $v_{pm}=\frac{2t_{p}}{%
\hbar }\sin (k_{m}^{p})$ the longitudinal velocity of the $m$th transverse
mode in lead $p$. The summation over the transverse coordinate $y_{q}$ runs
over from $1$ to $N_{q}$ ( $N_{q}$ is the width of lead $q$ )\cite{note2},
and the following orthogonality relations for transverse modes in lead $q$
have been applied in obtaining the last lines of Eq.(12a) and (12b): $%
\sum_{y_{q}}\chi _{m}^{q}(y_{q})\chi _{n}^{q}(y_{q})=\delta _{mn}$. It
should be noted that, if $p=q$, the results given by Eq.(12a) and (12b) will
denote actually the contribution of an incident electron from lead $q$ to
the charge and spin currents flowing in the same lead and $\phi _{qn\sigma
^{\prime }}^{pm\sigma }$ ( $\sigma ^{\prime }=\pm $ ) denote actually the
spin-flip ( $\sigma \neq \sigma ^{^{\prime }}$ ) and non-spin-flip ( $\sigma
=\sigma ^{^{\prime }}$ ) reflection amplitudes. ( See the explanations given
to Eqs.(3-4) in section II ). In such cases, the results given by Eq.(12a)
and (12b) can be expressed as the subtraction of the contributions due to
the incident wave ( i.e., the terms proportional to $\delta _{pq}$ in
Eq.(12a) and (12b) ) and the contributions due to the spin-flip and
non-spin-flip reflected waves.

The total terminal charge current $I_{q}$ and the total terminal spin
current $I_{q}^{\alpha }$ flowing in lead $q$ will be obtained by summing
the contributions of all incident electrons from all leads with the
corresponding density of states ( see the explanations given above Eq.(7) ).
Then we get that
\end{subequations}
\begin{subequations}
\begin{eqnarray}
I_{q} &=&\frac{e}{h}\sum_{pm\sigma }\int {d}Ef(E,\mu _{p})\verb|[|%
\sum_{n,\sigma ^{\prime }}|\phi _{qn\sigma ^{\prime }}^{pm\sigma }|^{2}\frac{%
v_{qn}}{v_{pm}}-\delta _{pq}\verb|]|  \notag \\
&=&\frac{e}{h}\sum_{p\sigma \sigma ^{\prime }}\int {d}Ef(E,\mu _{p})\verb|[|%
T_{q\sigma ^{\prime }}^{p\sigma }(E)-\delta _{pq}\delta _{\sigma \sigma
^{\prime }}N_{q}(E)\verb|]|  \notag \\
&=&\frac{e}{h}\sum_{p\sigma \sigma ^{\prime }}\int {d}E\verb|[|f(E,\mu
_{p})T_{q\sigma ^{\prime }}^{p\sigma }(E)-f(E,\mu _{q})T_{p\sigma ^{\prime
}}^{q\sigma }(E)\verb|]|,  \notag \\
&&
\end{eqnarray}%
\begin{eqnarray}
I_{q}^{\alpha } &=&\sum_{pm\sigma }\int {d}Ef(E,\mu _{p})[\sum_{n}\frac{%
v_{qn}}{4\pi v_{pm}}(|\phi _{qn+}^{pm\sigma }|^{2}-|\phi _{qn-}^{pm\sigma
}|^{2})]  \notag \\
&=&\frac{1}{4\pi }\sum_{p\sigma }\int {d}Ef(E,\mu _{p})[T_{q+}^{p\sigma
}(E)-T_{q-}^{p\sigma }(E)],
\end{eqnarray}%
where $T_{q\sigma ^{\prime }}^{p\sigma }(E)\equiv
\sum\limits_{m,n}\left\vert \phi _{qn\sigma ^{\prime }}^{pm\sigma
}\right\vert ^{2}\frac{v_{qn}}{v_{pm}}$ denotes ( by definition ) the
transmission probability from lead $p$ with spin $\sigma $ to lead $q$ with
spin $\sigma ^{\prime }$ ( see Ref.[33] and also the explanations given in
the appendix A ), and in obtaining the last line of Eq.(13a) the following
relation has been applied\cite{data}:
\end{subequations}
\begin{equation}
\sum_{p\sigma }T_{q\sigma ^{\prime }}^{p\sigma }(E)=\sum_{p\sigma }T_{q\bar{%
\sigma ^{\prime }}}^{p\sigma }(E)=N_{q}(E),
\end{equation}%
where $N_{q}(E)$ is the total number of conducting transverse modes in lead $%
q$ corresponding to a given energy $E$. \ As was discussed in detail in
Ref.[33], this relation follows directly from the unitarity of the S-matrix,
which is essential for the particle number conservation.

Eq.(13a) is just the usual Landauer-B\"{u}ttiker formula for terminal charge
currents in a multi-terminal mesoscopic system\cite{data}. The second line
in Eq.(13a) indicates clearly that the terminal charge current flowing in
lead $q$ can be expressed as the subtraction of the contributions due to all
incident modes ( corresponding to the terms proportional to $\delta _{pq}$ )
and the contributions due to all out-going modes ( corresponding to the
terms proportional to $T_{q\sigma ^{\prime }}^{p\sigma }$ ), which include
both the transmitted waves from other leads ( $p\neq q$ ) and the reflected
waves in lead $q$ ( $p=q$ ), noticing that $T_{q\sigma ^{\prime }}^{p\sigma }
$ denotes actually the spin-flip or non-spin-flip reflection probabilities
from lead $q$ to lead $q$ if $p=q$. Eq.(13b) is somewhat different from
Eq.(13a) in appearance, but the terminal spin current given by Eq.(13b) can
still be divided into two different kinds of contributions, namely the
contributions due to the transmitted waves from other leads ( corresponding
to those terms with $p\neq q$ in the summation $\sum_{p\sigma }[\ldots ]$ )
and the contributions due to the spin-flip and non-spin-flip reflected waves
in lead $q$ ( corresponding to those terms with $p=q$ in the summation $%
\sum_{p\sigma }[\ldots ]$ ).\cite{note3}

Eq.(13a) and (13b) are valid both in the equilibrium and in the
nonequilibrium states. In the equilibrium state, since $\mu _{p}\equiv \mu $
( independent of the lead label ) and $f(E,\mu _{p})\equiv f(E,\mu )$, in
Eq.(13a) and (13b) the summation $\sum_{p\sigma }[\ldots ]$ can be performed
first before carrying out the energy integration and we get that
\begin{subequations}
\begin{equation}
I_{q}=\frac{e}{h}\int {d}Ef(E,\mu )\sum_{p\sigma \sigma ^{\prime }}\verb|[|%
T_{q\sigma ^{\prime }}^{p\sigma }(E)-T_{p\sigma ^{\prime }}^{q\sigma }(E)%
\verb|]|,
\end{equation}%
\begin{equation}
I_{q}^{\alpha }=\frac{1}{4\pi }\int {d}Ef(E,\mu )\sum_{p\sigma
}[T_{q+}^{p\sigma }(E)-T_{q-}^{p\sigma }(E)].
\end{equation}%
Then by use of Eq.(14) one can see clearly that both terminal charge
currents and terminal spin currents will vanish exactly in the equilibrium
state. It should be stressed that in arriving at this conclusion we did not
involve the controversial issue of what is the correct definition of spin
current in the central SO coupled region at all, so those ambiguities that
might be caused by the use of an improper definition of spin current to the
SO coupled region have been eliminated in our derivations. Though we cannot
prove that the spin current also vanishes exactly inside the SO coupled
region based on the approach applied above, however, for a mesoscopic system
only the terminal ( charge or spin ) currents are the real useful quantities
from the $practical$ point of view ( i.e., one need to add external contacts
to induct the charge or spin currents out of a mesoscopic sample ). \ We
note that a similar conclusion as was obtained above has also been derived
in Ref.[37] based on some somewhat different arguments. Compared with the
derivations given in Ref.[37], the arguments given above seem to be more
simple and more transparent in principle. It also should be noted that,
based on a similar Landauer-B\"{u}ttiker formalism, it was argued in
Ref.[38] that the equilibrium terminal spin currents should indeed take
place in a three terminal system with spin-orbit coupling\cite{pareek}, in
contradiction to the conclusion obtained in the present paper and in
Ref.[37]. To our understandings, this contradiction was caused by the fact
that the contributions due to the spin-flip and non-spin-flip reflections in
the leads ( induced by the scatterings from the central SO coupled region )
was neglected in the calculations of terminal spin currents performed in
Ref.[38]. In contrast, in the calculations of terminal spin currents
performed in the present paper, the contributions due to the spin-flip and
non-spin-flip reflections in the leads induced by the scatterings from the
central SO coupled region have been treated in an accurate and strict way,
assuming that the leads are ideal and nonmagnetic.

\section{Some symmetry properties of nonequilibrium spin currents and spin
polarizations in two-terminal mesoscopic SO coupled systems}

When a multi-terminal mesoscopic SO coupled system is driven into a
nonequilibrium state ( i.e., there is charge current flow between different
leads ), nonequilibrium spin polarizations and/or terminal spin currents may
be induced by the charge current flow. In this section we discuss some
symmetry properties of such nonequilibrium spin currents and spin
polarizations. For clarity, we take a typical two-terminal mesoscopic
structure as shown in Fig.\ref{fig:twoterminal} as the example, where a
ballistic two-dimensional electron gas ( 2DEG ) with Rashba and/or k-linear
Dresselhaus SO coupling is attached to two ideal leads.
\begin{figure}[tbh]
\includegraphics[width=6cm,height=2cm]{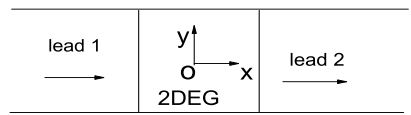}
\caption{Schematic geometry of a two-terminal mesoscopic SO coupled system.}
\end{figure}

\subsection{Non-antisymmetric lateral edge spin accumulations}

The study of nonequilibrium lateral edge spin accumulation induced by a
longitudinal charge current in a thin strip of a two-dimensional electron
gas with intrinsic SO coupling is of great theoretical interest because of
its close relations with the intrinsic spin Hall effect in such systems. It
was generally believed that the principal observable signature of the
intrinsic spin Hall effect in a SO coupled system is that, when a
longitudinal charge current circulates through such a system with a thin
strip geometry, \textit{antisymmetric} lateral edge spin accumulation (
polarized perpendicular to the 2DEG plane ) will be induced at the two
lateral edges of the strip due to the flow of the transverse spin Hall
current. Several recent numerical calculations had demonstrated that a
longitudinal charge current circulating through a thin strip of a ballistic
two-dimensional electron gas with Rashba SO coupling does can lead to the
generation of antisymmetric edge spin accumulation at the two lateral edges
of the strip, and the antisymmetric character of the transverse spatial
distribution of the lateral edge spin accumulation ( i.e., $\langle
S_{z}(x,y)\rangle =-\langle S_{z}(x,-y)\rangle $ ) had been argued to be a
strong support of the existence of intrinsic spin Hall effect in such
mesoscopic SO coupled systems\cite{NikolicPRB2005}. Here we discuss this
issue from a different point of view. We will show that, when a longitudinal
charge current circulates through a thin strip of a ballistic
two-dimensional electron gas with both Rashba and Dresselhaus SO coupling,
the transverse spatial distribution of the induced nonequilibrium lateral
edge spin accumulation ( polarized perpendicular to the 2DEG plane ) are in
general non-antisymmetric. The non-antisymmetric character of the lateral
edge spin accumulation contradicts seriously with the usual physical
pictures of spin Hall effect, though according to some theoretical
predictions, intrinsic spin Hall effect should also survive in the presence
of both Rashba and Dresselhaus SO coupling\cite{Sinitsyn2004}. The
non-antisymmetric character of the lateral edge spin accumulation implies
that, in addition to the intrinsic spin Hall effect, there may exist some
other physical reasons that might also lead to the generation of
nonequilibrium lateral edge spin accumulation in a SO coupled system ( with
a thin strip geometry ) when a longitudinal charge current circulates
through it.\cite{Usaj2005,JiangPRB2006}

Firstly let us look at what symmetry relations can be obtained for the
nonequilibrium lateral edge spin accumulation induced by a longitudinal
charge current based on the symmetry analysis of the Hamiltonian of the
system under study ( sketched in Fig.2 ). If only Rashba ( or only
Dresselhaus ) SO coupling presents, based on the symmetry properties of the
Hamiltonian of the structure under study, one can show rigorous that the
nonequilibrium lateral edge spin accumulation does should be antisymmetric
at the two lateral edges. Let us consider first the case in which only
Rashba SO coupling presents ( i.e., the Dresselhaus SO coupling strength is
zero ). If only Rashba SO coupling presents, from Eq.(2a) and Fig.2 one can
see that the Hamiltonian of the entire structure is invariant under the
combined transformation of the real space reflection $y\Rightarrow -y$ and
the spin space rotation around the $S_{y}$ axis ( with an angle $\pi $ ).
From this invariance one can get immediately that $\langle S_{z}(x,y)\rangle
_{I}=-\langle S_{z}(x,-y)\rangle _{I}$, where $\langle S_{z}\rangle _{I}$
denotes the nonequilibrium spin density induced by a longitudinal charge
current flowing from lead 1 to lead 2. It is interesting to note that this
antisymmetric relation can be deduced directly from the symmetry of the
structure under study but without need to resort to the concept of spin Hall
effect at all. Next, let us consider the case in which only Dresselhaus SO
coupling presents ( i.e., the Rashba SO coupling strength is zero ). If only
Dresselhaus SO coupling presents, then from Eq.(2b) and Fig.2 one can see
that the Hamiltonian of the entire structure is invariant under the combined
transformation of the real space reflection $y\Rightarrow -y$ and the spin
space rotation around the $S_{x}$ axis ( with an angle $\pi $ ). From this
invariance one also gets immediately that $\langle S_{z}(x,y)\rangle
_{I}=-\langle S_{z}(x,-y)\rangle _{I}$, i.e., the nonequilibrium lateral
edge spin accumulation still should be antisymmetric at the two lateral
edges if only Dresselhaus SO coupling presents.

\ If both Rashba and Dresselhaus SO couplings are present, then from
Eqs.(2a-2b) and Fig.2 one can see that the total Hamiltonian of the entire
structure is invariant under the combined transformation of the real space
center inversion $\mathbf{r}\Rightarrow -\mathbf{r}$ and the spin space
rotation around the $S_{z}$ axis ( with an angle $\pi $ ). From this
invariance one can get that $\langle S_{z}(x,y)\rangle _{I}=\langle
S_{z}(-x,-y)\rangle _{-I}$, where $\langle S_{z}\rangle _{-I}$ denotes the
nonequilibrium spin density induced by a longitudinal charge current flowing
from lead 2 to lead 1. On the other hand, from Eq.(\ref{eq:summationspin})
one can see that in the linear response regime one has
\end{subequations}
\begin{equation}
\langle S_{z}(x,y)\rangle _{I}=-\langle S_{z}(x,y)\rangle _{-I}.
\end{equation}%
Combining the two relations $\langle S_{z}(x,y)\rangle _{I}=\langle
S_{z}(-x,-y)\rangle _{-I}$ and $\langle S_{z}(x,y)\rangle _{I}=-\langle
S_{z}(x,y)\rangle _{-I}$, then the following symmetry relation can be
obtained for the nonequilibrium spin accumulation induced by a longitudinal
charge current flowing from lead 1 to lead 2: $\langle S_{z}(x,y)\rangle
_{I}=-\langle S_{z}(-x,-y)\rangle _{I}$. This symmetry relation implies that
the transverse spatial distribution of the nonequilibrium lateral edge spin
accumulation will be antisymmetric in the center cross-section of the strip,
i.e., $\langle S_{z}(0,y)\rangle _{I}=-\langle S_{z}(0,-y)\rangle _{I}$. Due
to the existence of this symmetry relation, one can deduce that in an
\textit{infinite} strip ( i.e, the length of the strip tends to infinity and
hence the effects of the contacted leads can be neglected ), the transverse
spatial distribution of the nonequilibrium lateral edge spin accumulation
will still be antisymmetric ( i.e., $\langle S_{z}(x,y)\rangle _{I}=-\langle
S_{z}(x,-y)\rangle _{I}$ for all $x$ ) in the presence of both Rashba and
Dresselhaus SO coupling. However, unlike the case in which only Rashba ( or
only Dresselhaus ) SO coupling presents, for a thin strip of finite length,
in the presence of both Rashba and Dresselhaus SO coupling, one cannot
deduce a general antisymmetric relation for the transverse spatial
distribution of the nonequilibrium lateral edge spin accumulation based on
the symmetry analysis of the total Hamiltonian of the entire structure under
study. From the theoretical points of view, this is due to the fact that, in
the presence of both Rashba and Dresselhaus SO coupling, the total
Hamiltonian of the entire structure under study ( i.e., a thin strip of
finite length contacted to two ideal leads ) is no longer invariant under
the combined transformation of the real space reflection $y\Rightarrow -y$
and the spin space rotation around the $S_{y}$ ( or $S_{x}$ ) axis with an
angle $\pi $. Indeed, as will be shown below, this non-antisymmetric
character can be verified by detailed numerical calculations.

One particular interesting case is that the Rashba and the Dresselhaus SO\
coupling strengths are equal ( i.e., $t_{R}=t_{D}$ or $t_{R}=-t_{D}$ ). In
this particular case, the total Hamiltonian is invariant under the following
unitary transformation in spin space ( while the real space coordinate $%
\mathbf{r}$ remain unchanged ): $\hat{U}_{+}\hat{H}\hat{U}_{+}=\hat{H}$ ( if
$t_{R}=t_{D}$ ) or $\hat{U}_{-}\hat{H}\hat{U}_{-}=\hat{H}$ ( if $t_{R}=-t_{D}
$ ), where $\hat{U}_{+}=(\hat{\sigma}_{x}+\hat{\sigma}_{y})/\sqrt{2}$ and $%
\hat{U}_{-}=(\hat{\sigma}_{x}-\hat{\sigma}_{y})/\sqrt{2}$. Under this
unitary transformation, the spin operators will transform as following: $%
\hat{\sigma}_{z}\rightarrow -\hat{\sigma}_{z}$, $\hat{\sigma}%
_{x}\rightleftharpoons \hat{\sigma}_{y}$ ( if $t_{R}=t_{D}$ ) or $\hat{\sigma%
}_{x}\rightleftharpoons -\hat{\sigma}_{y}$ ( if $t_{R}=-t_{D}$ ). Since the
real space coordinate $\mathbf{r}$ remain unchanged under this symmetry
manipulation, from the above symmetry properties in spin space one gets
immediately that $\langle S_{z}(x,y)\rangle _{I}=-\langle S_{z}(x,y)\rangle
_{I}$, suggesting that $\langle S_{z}(x,y)\rangle _{I}$ should vanish
everywhere if the Rashba and the Dresselhaus SO\ coupling strengths are
equal. This conclusion is in exact agreement with the corresponding
numerical results obtained based on the scattering wave approach introduced
in section II-III, which shows that $\langle S_{z}(x,y)\rangle _{I}$ does
vanish everywhere in the particular case of $t_{R}=t_{D}$ or $t_{R}=-t_{D}$.

To show more explicitly the non-antisymmetric character of the
lateral edge spin accumulation induced by a longitudinal charge
current in a 2DEG strip of finite length with both Rashba and
Dresselhaus SO coupling, in Fig.3 we plotted a typical pattern of
the two-dimensional spatial distribution of the nonequilibrium
spin density $\langle S_{z}\rangle $ in the strip obtained by
numerical calculations with the scattering wave function approach
introduced in Sec.II-III. In our numerical calculations we take
the typical values of the electron effective mass $m=0.04m_{e}$,
the lattice constant $a=3nm$, and the 2DEG strip contains
$120\times 40$ lattice sizes. The chemical potentials in the two
leads are set by fixing the longitudinal charge current to flow
from lead 1 to lead 2 as shown in Fig.2 and fixing the
longitudinal charge current density to $100\mu A/1.5\mu m$ ( as
reported in Ref.[26] ).
\begin{figure}[tbh]
\includegraphics[width=8cm,height=7cm]{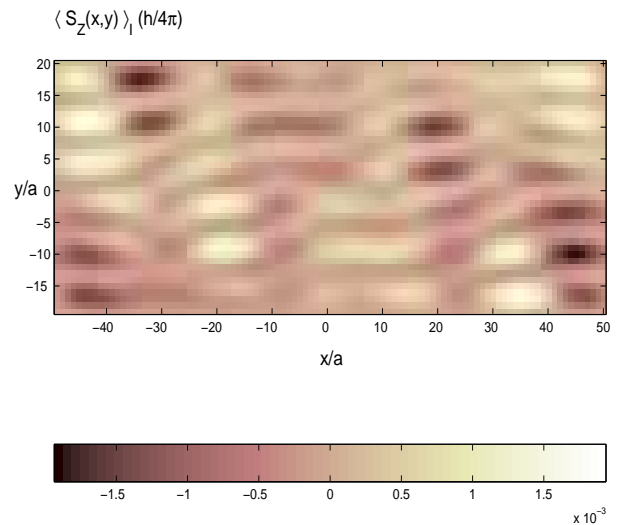}
\caption{(Color online)A typical pattern of the two-dimensional
spatial distribution of the current induced nonequilibrium spin
density $\langle S_{z}\rangle $ in a two-terminal structure (
sketched in Fig.2 ) in the presence of both Rashba and Dresselhaus
SO coupling. \ The Rashba and Dresselhaus SO coupling strength is
set to $t_{R}/t=0.08$ and $t_{D}/t=0.02$. }
\end{figure}
From Fig.3 one can see clearly that the transverse spatial distribution of
the nonequilibrium spin density $\langle S_{z}\rangle $ in the strip is
non-antisymmetric in general ( i.e., $\langle S_{z}(x,y)\rangle _{I}\neq
-\langle S_{z}(x,-y)\rangle _{I}$ for general $x$ ), except in the center
cross-section ( i.e., $x=0$ ) of the strip. The non-antisymmetric character
of the transverse spatial distribution of the nonequilibrium spin density $%
\langle S_{z}\rangle $ can be more clearly seen from Fig.4(a), where we
plotted several typical patterns of the profiles of the transverse spatial
distributions of the nonequilibrium spin density $\langle S_{z}\rangle $ in
a cross-section of the strip at $x\neq 0$. ( For comparison, the
corresponding results obtained in the case that only Rashba or only
Dresselhaus SO coupling presents were also plotted in Fig.4(b) ). The three
typical patterns shown in Fig.4(a) are obtained by fixing the Dresselhaus SO
coupling strength to $t_{D}=0.02t$ ( $t$ is the spin-independent hopping
parameter ) and varying the Rashba SO coupling strength $t_{R}$.
\begin{figure}[tbh]
\includegraphics[width=7cm,height=5cm]{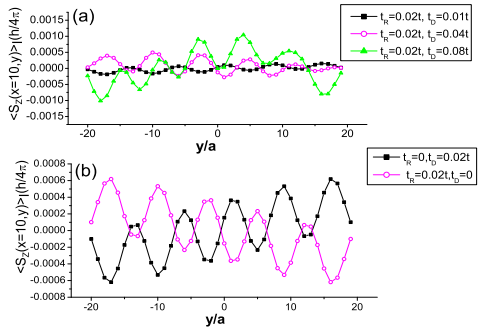}
\caption{(Color online)(a) Some typical profiles of the transverse
spatial distributions of the nonequilibrium spin density $\langle
S_{z}\rangle $ induced by a longitudinal charge current in a 2DEG
strip with both Rashba and Dresselhaus SO couplings, which shows
clearly that the transverse spatial distributions of $\langle
S_{z}\rangle $ are non-antisymmetric in general at both edges of
the strip in the presence of both Rashba and Dresselhaus SO
coupling. $\langle S_{z}\rangle $ vanishes everywhere in the
particular case of $t_{R}=t_{D}$ or $t_{R}=-t_{D}$ ( not shown
explicitly in the figure ). \ (b) The corresponding results
obtained in the case that only Rashba or only Dresselhaus SO
coupling presents, which shows clearly that the transverse spatial
distributions of $\langle S_{z}\rangle $ are antisymmetric at both
edges of the strip if only Rashba or only Dresselhaus SO coupling
presents. }
\end{figure}
From Fig.4(a) one can see clearly that, the transverse spatial distribution
of the nonequilibrium spin density $\langle S_{z}(x,y)\rangle _{I}$ can have
either the same signs or opposite signs at the two lateral edges of the
strip, depending on the ratios of $t_{R}/t_{D}$. Even in the case that $%
\langle S_{z}\rangle $ has opposite signs at the two lateral edges, the
transverse spatial distributions of $\langle S_{z}\rangle $ may still not be
antisymmetric ( i.e., $\langle S_{z}(x,y)\rangle \neq -\langle
S_{z}(x,-y)\rangle $ ), contradicting significantly with the usual physical
pictures of spin Hall effect. The non-antisymmetric character of the lateral
edge spin accumulation suggests that some cautions may need to be taken when
attributing the nonequilibrium lateral edge spin accumulation induced by a
longitudinal charge current in a thin strip of a SO coupled system to a spin
Hall effect, especially in the mesoscopic regime.

The results shown in Figs.3-4 are obtained in the absence of impurity
scatterings. One can show that the symmetry properties shown in Fig.3-4 are
robust against spinless weak impurity scatterings. To model spinless weak
disorder scatterings, we assume that the on-site energy at lattice sites in
the 2DEG strip are randomly distributed in a narrow energy region $[-W,W]$,
where $W$ is the amplitude of the on-site energy fluctuations characterizing
the disorder strength. ( In the absence of disorder scatterings, the on-site
energy at each lattice site can be set simply to zero. ) We calculate the
spin density for a number of random impurity configurations and then do
impurity average. In Fig.5 we show the variations of the transverse spatial
distributions of the nonequilibrium spin density $\langle S_{z}\rangle $ in
a cross-section of the strip as the disorder strength increases, from which
one can see that the symmetry properties of the transverse spatial
distributions of the nonequilibrium spin density are robust against spinless
weak disorder scatterings.
\begin{figure}[tbh]
\includegraphics[width=7cm,height=5cm]{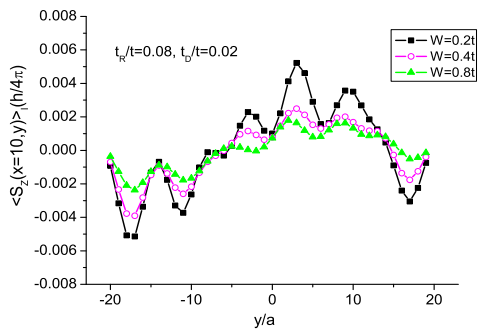}
\caption{(Color online)The profiles of the transverse spatial
distributions of the nonequilibrium spin density $\langle
S_{z}\rangle $ in the presence of disorder. We have done impurity
average over 1000 random impurity configurations for each case.}
\label{fig:disorder}
\end{figure}

\subsection{Non-conservative terminal spin currents}

As has been mentioned in the introduction section, a much controversial
issue related to the study of spin-polarized transports in SO coupled
systems is that whether spin currents should be a conserved quantity and
what is the correct definition of spin currents in such systems. It was now
well established that, in the presence of SO coupling, spin current
calculated based on the conventional definition is not a conserved quantity,
and in order to make spin current a conserved quantity in the presence of SO
coupling, significant modifications to the conventional definition will be
needed.\cite{Murakami04,shijunren} Nevertheless, it seemed that no consensus
had been arrived on whether spin current should be a conserved quantity in a
SO coupled system or whether there is a uniquely correct definition for spin
current in such a system.\cite{Murakami04,Sun05,shijunren,Jin05} Below we
will discuss this controversial issue from a different point of view, i.e.,
we do not consider the problem that what is the correct definition of spin
current in a SO coupled system but focus our discussion on the question that
whether the terminal spin currents in a multi-terminal mesoscopic SO coupled
system are conservative. As mentioned earlier, for a mesoscopic SO coupled
system, only the terminal spin currents are the real useful quantities. By
use of the two-probe mesoscopic structure shown in Fig.2 as the example, we
will show explicitly that the terminal spin currents in a multi-terminal
mesoscopic SO coupled system are non-conservative in general, i.e., the
total spin currents flowing into the SO coupled region are not equal to the
total spin currents flowing out of the same region. To illustrate this point
clearly, we take a two-terminal mesoscopic system with both Rashba and
Dresselhaus SO coupling as the example. In Fig.6(a) and (b) we plotted the
terminal spin currents $I_{1}^{z}$ and $I_{2}^{z}$ in the two leads ( with
spin parallel to the $z$ axis ) and the terminal spin currents $I_{1}^{y}$
and $I_{2}^{y}$ in the two leads ( with spin parallel to the $y$ axis ) as a
function of the Rashba SO coupling strength $t_{R}$, respectively. In our
calculations we fix the Dresselhaus SO coupling strength to $t_{D}=0.02t$
and fix the longitudinal charge current density to $100\mu A/1.5\mu m$. The
lattice constant $a=3nm$ and the lattice size of the 2DEG strip is taken to
be $100\times 40$. The positive direction of the spin current flow is
defined to be from lead $1$ to lead $2$.
\begin{figure}[tbh]
\includegraphics[width=7cm,height=7cm]{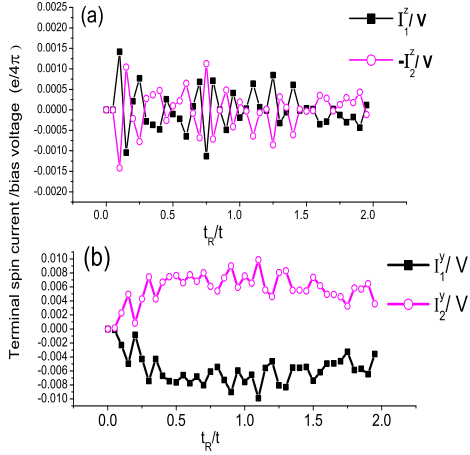}
\caption{(Color online)(a) The terminal spin currents $I_{1}^{z}$
and $-I_{2}^{z}$ (
divided by the voltage ) as a function of the Rashba SO coupling strenght $%
t_{R}$ ( in units of $t$ ). (b) The terminal spin currents $I_{1}^{y}$ and $%
I_{2}^{y}$ ( divided by the voltage ) as a function of the Rashba SO
coupling strenght $t_{R}$. The figures show that the terminal spin currents $%
I_{1}^{z}$ and $I_{2}^{z}$ in the two leads have the same signs and the
terminal spin currents $I_{1}^{y}$ and $I_{2}^{y}$ in the two leads have
opposite signs. ( Note that for clarity a minus sign is added before $%
I_{2}^{z}$ in Fig.6(a) ). The parameters used are given in the text or shown
in the figures. }
\label{fig:conductances}
\end{figure}
From Fig.6(a) one can see that the terminal spin currents $I_{1}^{z}$ and $%
I_{2}^{z}$ in the two leads have the same signs, which means that the
terminal spin currents with spin parallel to the $z$ axis will flow from
lead $1$ into the SO coupled region and then flow out of the SO coupled
region into lead $2$, similar to the usual charge current transport. From
Fig.6(b) one can see that the terminal spin currents $I_{1}^{y}$ and $%
I_{2}^{y}$ in the two leads have opposite signs, which means that the
terminal spin currents with spin parallel to the $y$ axis will flow out of
the SO coupled region in both leads and hence are non-conservative ( i.e.,
the spin current flowing into the SO coupled region does not equal to the
spin current flowing out of the same region ). Similarly one can show that
the terminal spin currents with spin parallel to the $x$ axis have also
opposite signs in the two leads, similar to the case shown in Fig.6(b). This
simple example illustrates explicitly that the terminal spin currents in a
multi-terminal mesoscopic SO coupled system are non-conservative in general.
It should be stressed that this non-conservation of terminal spin currents
is not caused by the use of an improper definition of spin current but is
intrinsic to spin-dependent transports in mesoscopic SO coupled systems. In
fact, in our calculations we did not involve the controversial issue of what
is the correct definition of spin current in the SO coupled region at all,
so the ambiguities that may be caused by the use of an improper definition
of spin current to the SO coupled region have been eliminated in our
calculations.

\section{\label{sec:4}Conclusion\newline
}

In summary, based on a scattering wave function approach, in this paper we
have studied theoretically some symmetry properties of spin currents and
spin polarizations in a multi-terminal mesoscopic structure in which a
spin-orbit coupled system is contacted to several ideal and nonmagnetic
external leads. Some interesting new results were obtained based on the
symmetry analysis of spin currents and spin polarizations in such a
multi-terminal mesoscopic structure. First, we showed that in the
equilibrium state no finite spin polarizations can exist both in the leads
and in the central SO coupled region and also no finite equilibrium terminal
spin currents can survive. Second, we showed that the lateral edge spin
accumulation induced by a longitudinal charge current in a thin strip of a
ballistic two-dimensional electron gas with both Rashba and Dresselhaus SO
coupling may be non-antisymmetric in general, which implies that some
cautions may need to be taken when attributing the nonequilibrium lateral
edge spin accumulation induced by a longitudinal charge current in a thin
strip of such a system to a spin Hall effect, especially in the mesoscopic
regime. Finally, by use of a typical two-probe structure as the example, we
showed explicitly that the nonequilibrium terminal spin currents in a
multi-terminal mesoscopic SO coupled system may be non-conservative in
general. Some symmetry properties discussed in the present paper might also
be helpful for clarifying some controversial issues related to the study of
spin-dependent transports in macroscopic SO coupled systems.

\begin{acknowledgments}
Y. J. Jiang was supported by the Natural Science Foundation of
Zhejiang province ( Grant No.Y605167 ). L. B. Hu was supported by
the National Science Foundation of China ( Grant No.10474022 ) and
the Natural Science Foundation of Guangdong province ( No.05200534
).
\end{acknowledgments}

\appendix

\section{Some details for the derivations of the scattering amplitudes and
the transmission probabilities}

In this appendix we give some details on how to derive the scattering
amplitudes and the transmission probabilities from the scattering wave
function approach introduced in Sec.II. For convenience of notation, we
arrange the scattering wave function $\psi ^{pm\sigma }(\mathbf{r}_{i})$
inside the SO coupled region into a column vector $\Psi _{s}$ whose
dimension is $2N$ ( $N$ is the total number of lattice sites in the SO
coupled region ) and arrange the scattering amplitudes $\phi _{qn\sigma
^{\prime }}^{pm\sigma }$ into a column vector $\Phi $ whose dimension is $2M$
( $M=\sum_{p}N_{p}$ and $N_{p}$ is the width of lead $p$ ). Substituting
Eqs.(3-4) into Eqs.(5a-5b) and making use of the orthogonality relations for
the transverse modes in the leads, one can show that the two column vectors $%
\Psi _{s}$ and $\Phi $ will satisfy the following relations:
\begin{equation}
\mathbf{A}\Psi _{s}=\mathbf{b}+\mathbf{B}\Phi ,\mathbf{C}\Phi =\mathbf{d}+%
\mathbf{D}\Psi _{s},  \label{eq:eqs1}
\end{equation}%
where $\mathbf{A},\mathbf{B},\mathbf{C},\mathbf{D}$ are four rectangular
matrices with the dimensions of $2N\times 2N$, $2N\times 2M$, $2M\times 2M$,
and $2M\times 2N$, respectively; $\mathbf{b}$\textbf{\ }and\textbf{\ }$%
\mathbf{d}$ are two column vectors with the dimensions of $2N$ and $2M$,
respectively. The elements of these matrices and column vectors can be
written down explicitly as
\begin{eqnarray}
\mathbf{A} &=&E\mathbf{I}-H_{sys},  \notag \\
\mathbf{B}(n_{p^{\prime \prime }y}\sigma ^{\prime \prime },p^{\prime
}m^{\prime }\sigma ^{\prime }) &=&-\delta _{p^{\prime \prime },p^{\prime
}}\delta _{\sigma ^{\prime \prime },\sigma ^{\prime }}t_{p^{\prime }s}\chi
_{m^{\prime }}^{p^{\prime }}(y_{p^{\prime \prime }})e^{ik_{m^{\prime
}}^{p^{\prime }}},  \notag \\
\mathbf{D}(p^{\prime }m^{\prime }\sigma ^{\prime },n_{p^{\prime \prime
}y}\sigma ^{\prime \prime }) &=&-\delta _{p^{\prime \prime },p^{\prime
}}\delta _{\sigma ^{\prime \prime },\sigma ^{\prime }}t_{p^{\prime }s}\chi
_{m^{\prime }}^{p^{\prime }}(y_{p^{\prime \prime }}),  \notag \\
\mathbf{C}(p^{\prime }m^{\prime }\sigma ^{\prime },p^{\prime \prime
}m^{\prime \prime }\sigma ^{\prime \prime }) &=&-\delta _{p^{\prime \prime
},p^{\prime }}\delta _{\sigma ^{\prime \prime },\sigma ^{\prime }}\delta
_{m^{\prime \prime }m^{\prime }}t_{p^{\prime }},  \notag \\
\mathbf{\ b}(n_{p^{\prime }y}\sigma ^{\prime }) &=&-\delta _{pp^{\prime
}}\delta _{\sigma \sigma ^{\prime }}t_{ps}\chi _{m}^{p}(y_{p^{\prime
}})e^{-ik_{m}^{p}},  \notag \\
\mathbf{d}(p^{\prime }m^{\prime }\sigma ^{\prime }) &=&\delta _{pp^{\prime
}}\delta _{mm^{\prime }}\delta _{\sigma \sigma ^{\prime }}t_{p},
\label{eq:matrix}
\end{eqnarray}%
where $\mathbf{I}$ stands for the identity matrix. The indices for leads,
transverse modes, lattice sites and spins can take all possible values. (
For simplicity of notation, we have used simply a symbol $n_{p^{\prime
}y^{\prime }}$ to denote a boundary lattice site in the SO coupled region
which is connected directly to a boundary lattice site $\mathbf{x}%
_{p^{\prime }}^{\prime }=(1,y_{p^{\prime }}^{\prime })$ in lead $p^{\prime }$%
. ) Eq.(\ref{eq:eqs1}) is just a compact form of the match conditions
(5a-5b) on the borders between the leads and the SO\ coupled region, from
which both the scattering amplitudes $\{\phi _{qn\sigma ^{\prime
}}^{pm\sigma }\}$ and the transmission probabilities $\{T_{q\sigma ^{\prime
}}^{p\sigma }\}$ can be obtained readily.

To derive a compact formula for the transmission probabilities between two
leads, we define an auxiliary matrix $\Sigma ^{R}\equiv \mathbf{B}\mathbf{C}%
^{-1}\mathbf{D}$. By use of Eq.(\ref{eq:matrix}) the matrix elements of $%
\Sigma ^{R}$ can be written down readily as following,
\begin{equation}
\Sigma ^{R}(n_{p^{\prime }y_{1}}\sigma ^{\prime },n_{p^{\prime }y_{2}}\sigma
^{\prime })=-\sum\limits_{m^{\prime }}\frac{t_{p^{\prime }s}^{2}}{%
t_{p^{\prime }}}\chi _{m^{\prime }}^{p^{\prime }}(y_{1})\chi _{m^{\prime
}}^{p^{\prime }}(y_{2})e^{ik_{m^{\prime }}^{p^{\prime }}},
\label{eq:selfenergy}
\end{equation}%
and all other matrix elements not shown explicitly above are zero. With the
help of this auxiliary matrix, from Eq.(\ref{eq:eqs1}) one can get that
\begin{equation}
\Psi _{s}=(\mathbf{A}-\Sigma ^{R})^{-1}(\mathbf{b}+\mathbf{B}\mathbf{C}^{-1}%
\mathbf{d})=G^{R}\mathbf{g,}  \label{eq:greenr}
\end{equation}%
where $\mathbf{g}$ is a column vector defined by $\mathbf{g}\equiv \mathbf{b}%
+\mathbf{B}\mathbf{C}^{-1}\mathbf{d}$ and $G^{R}$ is a matrix defined by $%
G^{R}\equiv (\mathbf{A}-\Sigma ^{R})^{-1}=[E\mathbf{I}-H_{sys}-\Sigma
^{R}]^{-1}$, which is just the usual retarded Green's function. By use of
Eq.(\ref{eq:matrix}), the elements of the column vector $\mathbf{g}$ can
also be written down readily as following,
\begin{equation}
\mathbf{g}(n_{p^{\prime }y}\sigma ^{\prime })=2i\delta _{pp^{\prime }}\delta
_{\sigma \sigma ^{\prime }}t_{ps}\sin (k_{m}^{p})\chi _{m}^{p}(y).
\end{equation}%
By substituting Eq.(\ref{eq:greenr}) into Eq.(\ref{eq:eqs1}) one gets that $%
\Phi =\mathbf{C}^{-1}\mathbf{d}+\mathbf{C}^{-1}\mathbf{D}G^{R}\mathbf{g}$.
Inserting Eq.(A5) into this formula and making use of Eq.(A2), one can show
readily that the scattering amplitudes $\phi _{qn\sigma ^{\prime
}}^{pm\sigma }$ will be given by
\begin{eqnarray}
\phi _{p^{\prime }m^{\prime }\sigma ^{\prime }}^{pm\sigma } &=&-\delta
_{pp^{\prime }}\delta _{mm^{\prime }}\delta _{\sigma \sigma ^{\prime
}}+2it_{p^{\prime }}^{-1}t_{p^{\prime }s}\sum_{y_{p},y_{p^{\prime
}}}t_{ps}\sin (k_{m}^{p})  \notag \\
&&\times \chi _{m^{\prime }}^{p^{\prime }}(y_{p^{\prime }})G_{\sigma
^{\prime }\sigma }^{R}(n_{y_{p^{\prime }}},n_{y_{p}})\chi _{m}(y_{p}).
\label{phi}
\end{eqnarray}%
The total transmission probability of a conduction electron from lead $p$ (
with spin index $\sigma $ ) to lead $p^{\prime }$ ( with spin index $\sigma
^{^{\prime }}$ ) is defined by $T_{p^{\prime }\sigma ^{\prime }}^{p\sigma
}=\sum\limits_{m,m^{\prime }}\left\vert \phi _{p^{\prime }m^{\prime }\sigma
^{\prime }}^{pm\sigma }\right\vert ^{2}\frac{v_{p^{\prime }m^{\prime }}}{%
v_{pm}}$, where $v_{p^{\prime }m^{\prime }}=\frac{1}{\hbar }2t_{p^{\prime
}}\sin (k_{m^{\prime }}^{p^{\prime }})$ is the longitudinal velocity of the
mode $m^{\prime }$ in lead $p^{\prime }$. Substituting Eq.(\ref{phi}) into
this formula, for $p\neq p^{\prime }$ one can get that
\begin{equation}
T_{p^{\prime }\sigma ^{\prime }}^{p\sigma }=Tr(\Gamma ^{p}G_{\sigma \sigma
^{\prime }}^{A}\Gamma ^{p^{\prime }}G_{\sigma ^{\prime }\sigma }^{R}),
\label{eq:LBformulaGreen}
\end{equation}%
where $G_{\sigma \sigma ^{\prime }}^{R}$ and $G_{\sigma \sigma ^{\prime
}}^{A}(\equiv G_{\sigma ^{\prime }\sigma }^{R\dagger })$ are the
spin-resolved retarded and advanced Green's functions, respectively, and $%
\Gamma ^{p}(y_{p},\overline{y}_{p})$ is defined by
\begin{equation}
\Gamma ^{p}(y_{p},\overline{y}_{p})=\sum\limits_{m}(\frac{t_{ps}}{t_{p}}%
)^{2}\chi _{m}(y_{p})v_{pm}\chi _{m}(\overline{y}_{p}).
\end{equation}%
The transmission probabilities given by Eq.(\ref{eq:LBformulaGreen}) have
exactly the same form as was obtained by the usual Green's function approach%
\cite{data}.


\begin{thebibliography}{99}
\bibitem{note1} When we calculate the terminal spin currents with spin
parallel to the $\alpha $ ( $=x,y,z$ ) axis, we will choose the $\alpha $
axis as the quantization axis of spins in the leads. This convention will be
assumed throughout the paper.
\bibitem{note2} The expectation values of $\hat{I}_{q}$ and $\hat{I}%
_{q}^{\alpha }$ given by Eq.(12a) and (12b) are independent of the
longitudinal coordinate $x_{q}$ in the steady state.
\bibitem{note3} Since we have assumed that the leads are ideal and
nonmagnetic, the contributions due to incident waves from lead $q$ with
opposite spin indexes ( given by the terms proportional to $\delta _{pq}$ in
Eq.(12b) ) cancel exactly with each other in Eq.(13b).
\end{thebibliography}
\end{document}